# Non-equilibrium evolution of the disoriented chiral condensate in heavy-ion collisions

Yuval Kluger

*Theoretical Division, Los Alamos National Laboratory*

*Los Alamos, New Mexico 87545 USA*

## ABSTRACT

We study the dynamics of the chiral phase transition expected during the expansion of the quark-gluon plasma produced in a high energy hadron or heavy ion collision in the $O(4)$ linear sigma model to leading order in a large $N$ expansion for strong coupling constants. Starting from an approximate equilibrium configuration at an initial proper time $\tau$ in the disordered phase, we study the transition to the ordered broken symmetry phase as the system expands and cools. We give results for the proper time evolution of the effective pion mass, the order parameter $<\sigma>$ as well as for the pion two point correlation function expressed in terms of a time dependent phase space number density and pair correlation density. We investigate the possibility of disoriented chiral condensate being formed during the expansion. In order to create large domains of disoriented chiral condensates low-momentum instabilities have to last for long enough periods of time. Our simulations show that instabilities that are formed during the initial stages of the expansion survive for proper times that are at most $3\,fm/c$.

## 1. Introduction

The possibility of producing large correlated regions of quark condensate $<\bar{q}_i q_j>$ pointing along the wrong direction in isospin space was proposed [1] to explain rare events where there is a deficit or excess of neutral pions observed in cosmic ray experiments [4]. The idea that such disoriented chiral condensates (DCC's) can be formed has been the source of several experimental proposals [2,5] in high energy collisions and a number of theoretical papers [3,6–12]. Theoretically, it has been recognized that the possibility of forming large regions of DCC relies on the existence of a substantially large regime in which the hot plasma formed after the collision evolves *out of equilibrium* [3]. In fact, if thermal equilibrium is approximately preserved by the dynamics, the typical correlation length would be determined by the pion mass and therefore would be too small to matter. For this reason, a number of authors made several attempts to analyze the nonequilibrium aspects of the dynamics of the chiral phase transition. These approaches vary in form and content: some authors performed numerical simulations on classical models [3,9,10], others used phenomenological terms – inspired by the classical kinetic theory – to model the interaction between the condensate and the quasiparticles [11] while some attempted to incorporate quantum and thermal fluctuations [6,12] into the picture.

There are clearly two important questions that the theoretical models should aim to answer. The first one is to determine whether during the evolution that follows the collision there are instabilities affecting the fluctuations, in which case there is a chance of the correlations growing. If this occurs, structure may tend to form through a process like spinodal decomposition.

The second question is, assuming the instability exists, can the correlated domains grow large enough so that many pions can be emitted from each domain making the detection of DCC's possible. The time scale for the instabilities to exist is strongly influenced by the strength of the interactions. In the strong coupling regime instabilities exist only for a short period of proper time. This makes it difficult to obtain large sizes for the correlated domains. On the other hand in a weakly coupled system the domains can grow for long periods of time, and a different picture emerges.

In our approach we do not put in instabilities by hand but look at typical fluctuations of initial conditions starting in a region of stability. Previous researchers have assumed the existence of an initial instability and focused their calculations on studying the growth of correlations. The typical and quite drastic way in which the instability had been previously introduced in the problem was by using the so called quench approximation. In a quench, the system (the hot plasma created after the collision) is subject to a sudden external action (the expansion into the vacuum) that has the following net effect: it does not change the state of the system but, instantaneously modifies the effective potential turning it "upside down". This is clearly an idealization that may be entirely inappropriate. In this paper we will present an approach that does not require this approximation and that enables us to study (in a simplified model) the existence of instabilities in a rather straightforward way.

Thus, our aim is to study the evolution of DCC's without imposing the quench approximation or any other ad hoc way of modeling the appearance of an instability. We start at an initial value of the proper time $\tau_0$ in the symmetric phase, where the particle masses are positive, by choosing a thermal distribution of particles above the critical temperature $T_c^2 = 12 f_\pi^2/N$ [20]. The system then cools by expanding into the vacuum. The expansion can be described by the Bjorken boost invariant picture for cooling [13] that has been applied in various hydrodynamic models to hadronic as well as heavy ion collisions.

The natural expansion and subsequent cooling of the plasma into the ordered vacuum is studied numerically by solving the update equations for the quantum modes as well as for the expectation values evolving in proper time. In this way, we are able to study the evolution of the plasma in a self consistent way, without imposing any instability by hand. We analyze various reasonable initial conditions on the fields, and determine whether they lead to instabilities. When the effective pion mass becomes negative instabilities ensue. We also determine the time evolving order parameter $<\sigma>$ and the spatial correlation function for the pion field.

In the investigations done so far, simple phenomenological models have been used hoping that they describe the fundamental physics involved in the dynamics of the phase transition. We will employ the linear sigma model, the most popular one in this context, which seems to have the essential attributes of being simple but realistic enough: it has the correct chiral symmetry properties and also appropriately describes the low energy phenomenology of pions. We will choose the parameters of the model to give reasonable values for three experimentally determined quantities : the mass of the pion $m_\pi$, the pion decay constant $f_\pi$ and the s wave $\pi - \pi$ phase shifts above threshold.

We remark that the theory only makes sense at cutoffs below the Landau pole. This limits the size of the renormalized coupling constant

$$\lambda_r^{max}(q^2 = 0) = \lim_{\lambda \to \infty} \frac{\lambda}{1 + N\lambda\Pi(\Lambda; q^2 = 0)} = \frac{2\pi^2}{\ln(\frac{2\Lambda}{m_\pi})}, \quad (1)$$

where $\lambda$ is the bare coupling and $\Pi$ is the polarization defined below. Since the mass difference between the $\sigma$ and $\pi$ is directly proportional to $\lambda_r(q^2 = m_{\sigma^2})$, this leads to an upper bound for the the mass of the $\sigma$ resonance as a function of $\Lambda$. Therefore, unlike at tree level, the mass of the $\sigma$ in the quantized theory is constrained in this model. $\Lambda$ is also constrained from the physical considerations that we want the mass of the $\sigma$ to be less than the cutoff. However, the mass of the $\sigma$ resonance increases as we decrease the cutoff since the renormalized coupling increases. This pins down the cutoff to lie between $0.7 GeV$ and $1 GeV$. For a cutoff of $1 GeV$ we find that $\lambda_r^{max}(q^2 = 0) \approx 11.5$ (taking into account finite cutoff effects). In our numerical simulations we use slightly smaller $\lambda_r$ in order to avoid numerical problems that occur in the vicinity of the Landau pole.

## 2. The linear $\sigma$-model in the large $N$ approximation

We will use the $O(4)$ linear sigma model to describe the evolution of the pions. We are well aware of the limitations of this approach, that provides a reasonable phenomenological model only for a limited range of energies (typically smaller than $1\ GeV$). In spite of its shortcomings, this model captures some of the essential physics involved in the dynamics of the phase transition that may produce disoriented chiral condensates. In particular, the chiral phase transition takes place at a reasonable temperature ($T_c = \sqrt{3}f_\pi$), and the low energy $\pi$ - $\pi$ scattering amplitudes are reasonable in this model. The mesons are organized in an $O(4)$ vector $\Phi = (\sigma, \vec{\pi})$ and the action is (in natural units $\hbar = c = 1$)

$$S = \int d^4x \{\frac{1}{2}\partial\Phi \cdot \partial\Phi - \frac{1}{4}\lambda(\Phi \cdot \Phi - v^2)^2 + H\sigma\}. \quad (2)$$

where we have use the Bjorken and Drell metric: $(1, -1, -1, -1)$. We will describe the evolution of the mean value $\bar{\Phi} \equiv <\Phi>$ and the two point correlation functions including the effects of quantum and thermal fluctuations. The complexity of the problem forces us to adopt some approximations. Perturbation theory is useless for our purpose [14,15], and a scheme which is non–perturbative in $\lambda$ must be adopted. In this context, the most popular approaches which allow for a real time analysis are the the Hartree (or Gaussian) ansatz [16] and the large $N$ expansion of the $O(N)$ sigma model [17]. We will adopt the latter since it presents several advantages. On the one hand, the expansion is systematic and allows us to study higher order corrections (work is in progress in this direction and results will be presented elsewhere [18]), and on the other hand, when using the Hartree ansatz in the context of the study of DCC's, one is forced to take the large $N$ limit as well(see Ref. [12]). This is due to the well known fact that the Gaussian approximation violates the Goldstone theorem giving an unphysical (and not necessarily small) mass to the pions in the $H = 0$ limit. Of course, the approximation we adopt here is not expected to capture all the features of the phase transition (as is well known, mean field theory fails to predict the correct critical exponents, but allows us to explore the strong coupling regime).

The large $N$ effective equations can be obtained in a variety of ways, which are extensively discussed in the literature [17]. A very convenient method is to use an alternative equivalent action, which is a functional of the original fields $\Phi$ and of the auxiliary constrained field $\chi = \lambda(\Phi^2/2N - v^2)$ [19].

$$\tilde{S}[\Phi, \chi] = \int d^4x \{-\frac{1}{2}\phi_i(\Box + \chi)\phi_i + \frac{\chi^2}{4\lambda} + \frac{1}{2}\chi v^2 + H\sigma\}. \tag{3}$$

As this action is now quadratic in $\Phi$ we can perform the functional integral over those fields, and are left with a functional integration over $\chi$ which, to leading order in $1/N$ can be calculated by the stationary phase method. The effective action, which is a functional of the mean fields of $\Phi$ and $\chi$, is obtained by Legandre transformation of the generating functional of the connected Green's functions. To the lowest order (in large $N$ expansion) the effective action is given by

$$\Gamma[\Phi, \chi] = \int d^4x \{-\frac{1}{2}\phi_i(\Box + \chi)\phi_i + \frac{\chi^2}{4\lambda} + \frac{1}{2}\chi v^2 + H\sigma + \frac{i}{2}N\mathrm{tr}\,\ln G_0^{-1}\}, \tag{4}$$

where

$$G_0^{-1}(x, y) = i[\Box + \chi(x)]\,\delta^4(x - y), \tag{5}$$

and where for notational convenience we drop the overbars and denote the expectation values as $\Phi$ and $\chi$.

From this effective action we derive the equations of motion for the fields $\Phi$ and the equation of constraint (gap equation) for the composite field $\chi$ which plays the role of the effective mass for the mean values $\Phi_i$:

$$(\Box_x + \chi(x))\Phi_i(x) = H\delta_{i1} \tag{6}$$

$$\chi(x) = \lambda(-v^2 + \Phi^2(x) + NG_0(x,x)). \tag{7}$$

The function $G_0(x, x)$ that appears in (7) is the coincidence limit of the propagator $G_0(x, y)$ that inverts the operator $G_0^{-1}$ defined in (5). We can use an auxiliary quantum field $\phi(x)$ where $<\phi(x)>= 0$ and

$$(\Box_x + \chi(x))\phi(x) = 0 \tag{8}$$

to construct this propagator as $G_0(x, y) =< T \phi(x)\phi(y) >$.

The initial value problem associated with equations (5)–(7) will be solved in the next section. Here, we would like to address the issue of how to use this model to make contact with the phenomenology we want to describe. Thus, we must fix the values of the parameters appearing in the above equations so that they describe low energy pion physics. The measurable quantities we want to reproduce are the pion mass $m_\pi = 135\ MeV$, the pion decay constant $f_\pi = 92.5\ MeV$ as well as the s wave, $I = 0$ phase shifts in the energy range $300 - 420 MeV$. To fit these physical quantities we analyze our equations in the "true vacuum" state (i.e., in equilibrium at zero temperature). In this state the derivatives of the expectation values vanish, and we have $\Phi = (\sigma_v, \vec{0})$, $\chi = \chi_v$ where $\sigma_v$ and $\chi_v$ are some constants whose values will be determined below. The physical masses can be related to the parameters of the theory by computing the inverse propagators of the pion and sigma fields. The s-wave phase shifts are determined from the $\pi - \pi$ scattering amplitudes obtained in this approximation which are controlled by the exchange of the composite field $\chi$ propagator. In the vacuum sector (where $\chi = m_\pi^2$, $\pi_i = 0$ and $\sigma = f_\pi$) the $\pi$, $\sigma$, and $\chi$ propagators are obtained by inverting the matrix inverse propagator

$$\hat{G}^{-1} = \frac{\delta^2 \Gamma}{\delta \Psi \delta \Psi}, \tag{9}$$

with the 5 dimensional vector $\Psi \equiv (\Phi, \chi)$, and are given by

$$\hat{G}_{\pi\pi}(p^2)_{ij} = \left[\delta_{ij}(p^2 - \chi) - \pi_i \hat{G}_{0\ \chi\chi}(p^2)\pi_j\right]^{-1}, \tag{10}$$

$$\hat{G}_{\sigma\sigma}(p^2) = \left[p^2 - m^2 - \sigma^2 \hat{G}_{0\ \chi\chi}(p^2)\right]^{-1}, \tag{11}$$

$$\hat{G}_{\chi\chi}(p^2) = \left[G_0^{-1}{}_{\chi\chi}(p^2) + \frac{\sigma^2}{\chi - p^2}\right]^{-1}, \tag{12}$$

where

$$\hat{G}_0^{-1}{}_{\chi\chi}(p^2) = \frac{1}{2\lambda} + \frac{N}{2}\Pi(p^2),$$

$$\Pi(p^2) = iG_0^2 = -i\int [d^4q]\ (\chi - q^2)^{-1}(\chi - (p+q)^2)^{-1}. \tag{13}$$

We use the notation that $[d^n q] = d^n q/(2\pi)^n$. (Notice that in the initial value problem one can have $\pi_i \neq 0$. Examining Eq. (10) we realize that $\delta_{ij}G_0(x,y)$ is the pion propagator only when $<\pi_i(x)>= 0$.)

The pion-pion scattering invariant T-matrix is given by:

$$T = \delta_{ij}\delta_{kl}A(s) + \delta_{ik}\delta_{jl}A(t) + \delta_{il}\delta_{jk}A(u), \tag{14}$$

where the isospin indices $i, j, k, l$ are coupled to the four momenta $p_1, p_2, p_3, p_4$ such that $s = (p_1 + p_2)^2$, $t = (p_1 - p_3)^2$ $u = (p_1 - p_4)^2$. In leading order in large $N$ the amplitudes $A(s)$, $A(t)$ and $A(u)$ are exactly the $\chi$ propagator, namely

$$A(s) = -\hat{G}_{\chi\chi}(p^2 = s). \tag{15}$$

The large N expansion preserves the current algebra so that the usual low energy theorem is exact. That is, for small $s, m^2$ one easily shows that

$$A(s) \to \frac{(s - m^2)}{f_\pi^2}.$$

This amplitude is independent of the coupling constant $\lambda$. We thus need to go above threshold to determine the coupling constant. The $I = 0$ scattering amplitude is

$$A^0 = 3A(s) + A(t) + A(u). \tag{16}$$

The s-wave scattering amplitude is obtained by integrating the the I=0 scattering amplitude over angles.

$$f_{l=0} = e^{i\delta(s)}\sin\delta(s) = \frac{1}{16\pi}\sqrt{1 - \frac{4m^2}{s}}\frac{1}{2}\int_{-1}^{1} dz\ A^0, \tag{17}$$

where $z = \cos\theta$ and $\theta$ is the scattering angle in the s channel center of mass system.

The vacuum expectation value of $\sigma$ is determined by $f_\pi$

$$\sigma_v = f_\pi. \tag{18}$$

On the other hand, in the vacuum we have $\vec{\pi} = 0$, and the pion inverse propagator is $G^{-1}_{ij\ \pi,\pi}(x,y) = G_0^{-1}(x,y)\delta_{ij}$. Therefore, the vacuum expectation value of $\chi$ is

$$\chi_v = m_\pi^2 \equiv m^2. \tag{19}$$

The sigma mass can be approximately determined in terms of the inverse sigma propagator as the zero in the real part of the inverse propagator (a more precise determination which gives a slightly different result is from the peak in the $I = 0$, $l = 0$ scattering amplitude). This leads to the equation:

$$m_\sigma^2 = m^2 + \sigma^2 Re[G_{0\ \chi\chi}(m_\sigma^2)]. \tag{20}$$

We do not expect the theory to be valid for energies above 1 GeV since in that regime the correct dynamics is described by QCD. In fact if we raise $\Lambda$ above 1 GeV, the maximum value of the renormalized coupling goes down, and the $\sigma$ mass becomes unacceptably low. Reasonable values for the mass of the $\sigma$ constrain $\Lambda$ to be in the range $0.7 GeV < \Lambda \leq 1 GeV$. In that range the best value of the renormalized coupling $\lambda_r$ is between 7-10. For this range of values for $\lambda_r$ this model agrees qualitatively with low energy scattering data as was mentioned before. With this choice of parameters, it is difficult to obtain values of the $\sigma$ mass higher than $450 MeV$.

## 3. Cooling by expansion

To analyze the possibility of forming DCC's we should take into account the specific features characterizing the situation after a highly energetic heavy ion collision. Experimentally, a flat plateau in the distribution of produced particles per unit rapidity is observed in the central rapidity region (these results are obtained in $p$–$\bar{p}$ and other collisions). This suggests the existence of an approximate Lorentz boost invariance. Thus, the simplest picture of a collision, due to Landau [25], is one in which the excited nuclei are highly contracted pancakes receding away from the collision point at approximately the speed of light. The boost invariance implies that the evolution of the "hot plasma" that is left in between the nuclei looks the same when viewed from different inertial frames. Of course, this is an approximate picture that is not valid for large values of the spatial rapidity and for transverse distances of the order of the nucleus size. The existence of this approximate symmetry can be used to make a very simple hydrodynamical model [13] that, in some cases, may describe the evolution of the plasma. It is worth reviewing these ideas very briefly. Firstly, one should recognize that the natural coordinates to make a boost invariant model are the proper time $\tau$ and the spatial rapidity $\eta$, defined as

$$\tau \equiv (t^2 - x^2)^{1/2}, \qquad \eta \equiv \frac{1}{2} \log(\frac{t-x}{t+x}). \tag{21}$$

The observed symmetry will be respected by the model if one imposes initial values on a $\tau =$constant hypersurface (and not at constant laboratory time $t$).

According to this simple model of a collision, the plasma evolves in a highly inhomogeneous way when viewed from the laboratory frame. In fact, by analyzing a constant $t$ surface we realize that the field configurations strongly depend on the spatial coordinate $x$. Near the light cone $|x| = t$ the system is "hot" (corresponding to small values of the proper time $\tau$). On the other hand, for small values of $x$ (that correspond to larger values of $\tau$) the system is "colder". This type of configuration, hot in the outside and cold in the inside, is schematically known as "Baked Alaska".

In this paper we want to study the formation of DCC's using some of the ideas presented above. We *will not* assume a quasi equilibrium situation or use a phenomenological hydrodynamical (or kinetic) model to describe the evolution of our system. On the contrary, we will study the nonequilibrium evolution in its full glory and solve equations (5)–(7), which include thermal and quantum effects. Using the coordinates (21) and fixing boost invariant initial conditions at an initial proper time $\tau_0$ we introduce the expansion and "cooling" of the plasma in a natural way. Therefore, we do not need to introduce any ad hoc cooling mechanisms by hand. The cooling, if any, will appear as a result of the evolution, which is fully out of equilibrium (a similar approach in different kind of expansion has been investigated independently by Gavin and Mueller[11]). In this way, we can really test the validity of the "quenching" approximation that has been almost universally used when analyzing the evolution of DCC's.

### 3.1. The equations

In the coordinates (21) Minkowsky's arc element is

$$ds^2 = d\tau^2 - \tau^2 d\eta^2 - dx_\perp^2. \tag{22}$$

To study the evolution of the mean fields and correlations in these coordinates, our first task is to rewrite equations (5)–(7) using the new variables. Assuming that the mean values $\Phi$ and $\chi$ are functions of $\tau$ only (homogeneity in the constant $\tau$ hypersurface) we have

$$\tau^{-1}\partial_\tau\, \tau\partial_\tau\, \Phi_i(\tau) + \chi(\tau)\, \Phi_i(\tau) = H\delta_{i1} \tag{23}$$

$$\chi(\tau) = \lambda(-v^2 + \Phi_i^2(\tau) + N < \phi^2(x,\tau) >), \tag{24}$$

where the quantum field $\phi(x,\tau)$ satisfies the Klein Gordon equation

$$\left(\tau^{-1}\partial_\tau\, \tau\partial_\tau\, - \tau^{-2}\partial_\eta^2 - \partial_\perp^2 + \chi(x)\right)\phi(x,\tau) = 0. \tag{25}$$

The quantum field $\phi(x,\tau)$ defined here is an auxiliary field which allows us to calculate the Wightman function $G_0(x,y)$ by taking the expectation value:

$$G_0(x,y;\tau) \equiv < \phi(x,\tau)\, \phi(y,\tau) > . \tag{26}$$

When the expectation value of the pion field is zero, then this field corresponds to one component of the pion field.

We expand this field in an orthonormal basis

$$\phi(\eta, x_\perp, \tau) \equiv \frac{1}{\tau^{1/2}} \int [d^3\mathrm{k}](\exp(i\mathrm{k}\cdot\mathbf{x})f_\mathrm{k}(\tau)\, a_\mathrm{k} + h.c.), \tag{27}$$

where $\mathrm{k}\cdot\mathbf{x} \equiv k_\eta\eta + \vec{k}_\perp \vec{x}_\perp$, $[d^3\mathrm{k}] \equiv dk_\eta d^2k_\perp/(2\pi)^3$ and the mode functions $f_\mathrm{k}(\tau)$ evolve according to:

$$\ddot{f}_\mathrm{k} + (\frac{k_\eta^2}{\tau^2} + \vec{k}_\perp^2 + \chi(\tau) + \frac{1}{4\tau^2})f_\mathrm{k} = 0. \tag{28}$$

A dot here denotes the derivative with respect to the proper time $\tau$. The expectation value $<\phi^2(x,\tau)>$ can be expressed in terms of the mode functions $f_k$ and of the distribution functions

$$n_{\mathrm{k}} \equiv < a_{\mathrm{k}}^\dagger a_{\mathrm{k}} >, \quad g_{\mathrm{k}} \equiv < a_{\mathrm{k}} a_{\mathrm{k}} >, \tag{29}$$

which entirely characterize the initial state of the quantum field. For simplicity, we will assume that the initial state is described by a density matrix which is diagonal in the number basis (like a thermal state). In such a case, the only nonvanishing distribution is $n_{\mathrm{k}}$. Thus, replacing the above expressions in (24) we have:

$$\chi(\tau) = \lambda\Big(-v^2 + \Phi_i^2(\tau) + \frac{1}{\tau} N \int [d^3 \mathrm{k}] |f_{\mathrm{k}}(\tau)|^2 \, (1 + 2\, n_{\mathrm{k}})\Big). \tag{30}$$

We assume that the initial density matrix at $\tau_0$ is one of local thermal equilibrium in the comoving frame

$$n_{\mathrm{k}}(\tau_0) \equiv n_{\mathrm{k}} = \frac{1}{e^{\beta_0 E_{\mathrm{k}}^0} - 1}, \tag{31}$$

where $\beta_0 = 1/T_0$, $E_{\mathrm{k}}^0 = \sqrt{k_\eta^2/\tau^2 + k_\perp^2 + \chi(\tau_0)}$.

To renormalize the theory we absorb the quadratic divergences in the bare mass $\lambda v^2$ and the logarithmic ones in the coupling constant $\lambda$. After a few simple manipulations (that involve adding and subtracting the appropriate terms in (30)) we can write the equation for $\chi$ as

$$\chi(\tau) = \chi(\tau_0) \; + \; \frac{\lambda_r}{Z}(\Phi_i^2(\tau) - \Phi_i^2(\tau_0)) + \frac{N\lambda_r}{\tau Z} \int [d^3\mathrm{k}]\Big\{|f_{\mathrm{k}}(\tau)|^2(1 + 2\, n_{\mathrm{k}}) -$$
$$- \; \frac{1}{2\tilde{\omega}_{\mathrm{k}}(\tau)}\Big\} - \frac{N\lambda_r}{\tau_0 Z}\int [d^3\mathrm{k}]\frac{1}{\omega_{\mathrm{k}}(\tau_0)}n_{\mathrm{k}}, \tag{32}$$

where $\tilde{\omega}_{\mathrm{k}}(\tau) \equiv (k_\eta^2/\tau^2 + \vec{k}_\perp^2 + \chi(\tau_0))^{1/2}$ and the renormalized coupling $\lambda_r$ and $Z$ are defined as

$$\lambda = \lambda_r/Z, \qquad Z = 1 - \lambda_r \delta\lambda, \qquad \delta\lambda \equiv \frac{N}{4\tau}\int^\Lambda [d^3\mathrm{k}]\frac{1}{\tilde{\omega}_{\mathrm{k}}^3(\tau)}. \tag{33}$$

The initial value $\chi(\tau_0)$ comes from solving the gap equation (30) at the initial time using the fact that in the initial equilibrium state at $\tau_0$, $\Phi_{eq}(T_0) = H\delta_{i1}/\chi_{eq}(T_0)$. Notice that the set of coupled equations, (23), (28), (32), are entirely written in terms of the renormalized quantities $\lambda_r$ and $\chi_0$. Changing the value of the cutoff will change both the value of the integrals and the value of $Z$ in Eq. (32) leaving $\chi$ cutoff independent.

### 3.2. The correlation function

Expanding the field $\phi$ in terms of the first order adiabatic mode functions,

$$f_k^0 = \frac{e^{-i\int_{\tau_0}^{\tau} d\tau' \omega_k(\tau')}}{\sqrt{2\omega_k(\tau)}}, \tag{34}$$

where $\omega_k(\tau) \equiv (\frac{k_\eta^2 + 1/4}{\tau^2} + \vec{k}_\perp^2 + \chi(\tau))^{1/2}$, we have

$$\phi(\eta, x_\perp, \tau) \equiv \frac{1}{\tau^{1/2}} \int [d^3k](\exp(i k \cdot x) f_k^0(\tau)\ a_k(\tau)\ + h.c.). \tag{35}$$

The two sets of creation and annihilation operators are connected by a Bogoliubov transformation:

$$a_k(\tau) = \alpha(k, \tau) a_k + \beta(k, \tau) a_{-k}^\dagger, \tag{36}$$

where the Bogoliubov coeficients are given by

$$\alpha(k, \tau) = i(f_k^{0*} \frac{\partial f_k}{\partial \tau} - \frac{\partial f_k^{0*}}{\partial \tau} f_k) \tag{37}$$

$$\beta(k, \tau) = i(f_k^0 \frac{\partial f_k}{\partial \tau} - \frac{\partial f_k^0}{\partial \tau} f_k). \tag{38}$$

The time dependent creation and annihilation operators satisfy

$$\dot{a}_k f_k^0 + \dot{a}_k^\dagger f_k^{*0} = 0 \tag{39}$$

in order for the usual canonical commutation relations to hold. Then we can write the Green function $G^0$ as follows:

$$\begin{aligned} G^0(x, y; \tau) &= \frac{1}{\tau} \int [d^3k] e^{ik\cdot(x-y)} |f_k(\tau)|^2\ (1 + 2\ n_k(\tau_0)) \\ &= \frac{1}{\tau} \int [d^3k] e^{ik\cdot(x-y)} \frac{(1 + 2\ n_k(\tau) + 2Re[g_k(\tau) e^{-2i\int_{\tau_0}^{\tau} d\tau' \omega_k(\tau')}])}{2\omega_k(\tau)}, \end{aligned} \tag{40}$$

where

$$\begin{aligned} <a_k^\dagger(\tau) a_q(\tau)> &= (2\pi)^3 \delta^3(k-q) n_k(\tau) \\ <a_{-k}(\tau) a_q(\tau)> &= (2\pi)^3 \delta^3(k-q) g_k(\tau). \end{aligned} \tag{41}$$

In terms of the initial distribution of particles $n_k(\tau_0)$ and the Bogoliubov coefficients $\alpha$ and $\beta$ we have:

$$n_k(\tau) = n_k(\tau_0) + |\beta(k, \tau)|^2 (1 + 2n_k(\tau_0)) \tag{42}$$

$$g_{\mathrm{k}}(\tau) = \alpha(\mathrm{k},\tau)\beta(\mathrm{k},\tau)(1 + 2n_{\mathrm{k}}(\tau_0)). \tag{43}$$

The variables $n(\tau)$ and $g(\tau)$ now have the physical interpretation as the interpolating phase space number and pair density for the case where $<\pi_i>=0$. These operators coincide with the time independent number and pair densities defined by (29) at $\tau_0$, and in the out regime become the physically measurable number and pair densities. For $<\pi_i>\neq 0$ the actual pion two point function is more complicated than $G^0$ and this is only one piece of that Green's function. Of course in the vacuum Isospin conservation requires $<\pi_i>=0$, so that once we are approaching the final true vacuum state during the cooling process, one can use these interpolating number and pair operators to describe the physics of the problem. Also note that during the evolution, $\omega_k^2$ can become negative (the instability phase) and for those modes the adiabatic basis does not exist.

## 4. Initial conditions and results

In order to solve equations (23), (28) and (30) as an initial value problem, we need to give Cauchy data (the function and its derivatives) for the mean values $\Phi_i(\tau)$ and for the mode functions $f_k(\tau)$. It is possible to choose the initial data $f_k(\tau_0)$ and $\dot{f}_k(\tau_0)$ so that the vacuum state coincides with the ordinary Minkowsky vacuum, at least for high momentum. It can be shown that this is accomplished by taking the "zero order adiabatic" vacuum [26] where

$$f_{\mathrm{k}}(\tau_0) = \frac{1}{2\omega_{\mathrm{k}}(\tau_0)}, \qquad \dot{f}_{\mathrm{k}}(\tau_0) = \left(i\,\omega_{\mathrm{k}}(\tau_0) - \frac{\dot{\omega}_{\mathrm{k}}(\tau_0)}{2\omega_{\mathrm{k}}(\tau_0)}\right)f_{\mathrm{k}}(\tau_0), \tag{44}$$

where $\omega_{\mathrm{k}}(\tau) \equiv (k_\eta^2/\tau^2 + \vec{k}_\perp^2 + \chi(\tau))^{1/2}$.

We should point out that the initial value of $\chi$ will always be restricted to being positive. As we said, $\chi(\tau)$ is the effective mass of the quasiparticles. Therefore, taking a negative initial value for $\chi$ implies that we are "turning the effective potential upside down." In our view, this cannot be the consequence of forming a "peculiar" initial state but must rather be the consequence of the cooling mechanism, which is entirely produced by the expansion. In fact, starting with a negative $\chi_0$ is what is done when studying this problem by using the quench approximation: one starts with a hot initial state and lets it evolve in the low temperature effective potential. It should be clear by now that this is drastically different from our approach. We will study the selfconsistent evolution of $\chi$ starting from a "hot" initial value and follow its evolution.

We want to start in the quark - gluon plasma phase above the chiral phase transition which places constraints on the initial energy density present at $\tau_0$. We then assume that slightly above this transition it is reasonable to model the chiral transition with the effective Lagrangian of the sigma model. Not knowing exactly what this proper time is, we will consider here reasonable initial proper times ($1fm/c < \tau_0 < 4fm/c$) and study the effect of the initial proper time $\tau$ on the production of instabilities. Taking larger proper times as the starting point for our calculation reduces even further the possibility of instability growth.

The first issue we will examine here is the existence of instabilities. Notice that there are unstable modes $f_k$ whenever $(k_\eta^2 + 1/4)/\tau^2 + k_\perp^2 + \chi$ is negative so that only the long wave length modes can become unstable. $\chi$ can classically become negative only when

$$\chi(\tau_0) + \lambda\Phi_i^2(\tau) - \lambda\Phi_i^2(\tau_0) < 0, \tag{45}$$

so that the most unstable case has $\Phi(\tau) = 0$. Therefore, at the classical level the bare coupling must satisfy

$$\lambda > \frac{\chi(\tau_0)}{\Phi_i^2(\tau_0)} \tag{46}$$

for there to be any instabilities. Once instabilities grow, they cause exponential growth of the modes in the mode sum which contributes a positive quantity to the equation for $\chi$. Thus, the stronger the coupling the quicker $\chi$ returns to a positive value.

In previous works the presence of unstable modes was assumed by imposing the quench approximation. In our numerical studies we find that, due to the strong coupling, in a wide class of initial conditions consistent with reasonable fluctuations found in a thermal distribution no instabilities develop. In our simulations the one thing we did not change was the value of the composite field $\chi$ which was fixed to be the solution of the gap equation in the initial thermal state. We also maintained the constraint that the initial values of the expectation values of the field, namely $\pi(\tau_0)$ and $\sigma(\tau_0)$ satisfied the relationship:

$$\vec{\pi}^2(\tau_0) + \sigma^2(\tau_0) = \sigma_T^2,$$

where $\sigma_T$ is the equilibrium value of $\Phi$ at the initial temperature $T$. This last constraint is reasonable since it doe not cause much energy to make a rotation in the direction of $\pi_i$. Thus we chose different initial $\sigma$ and $\pi_i$ consistent with the above constraints and then probed the effect of different initial values for $\dot\sigma$ and $\dot\pi_i$. Since the results for $\dot\pi_i \neq 0$ were similar to those when $\dot\sigma \neq 0$ we mainly present here the results for the latter case. We have also surveyed other possibilities that violate the above constraints such as choosing initial $\sigma = 0$. In this case we found that instabilities (even with $\dot\sigma \neq 0$) have shorter life times than those with the above constraint.

We have performed numerical simulations on the connection machine CM-5 using a grid that has 100 modes in the transverse direction and $100(\tau/\tau_0)$ modes in the $\eta$ direction with $dk_\eta/fm = \Lambda/100$, $dk_\perp = \Lambda/100$.

First, let us consider the case where the initial value $\sigma(\tau_0)$ is the thermal equilibrium one corresponding to a temperature of $200 MeV$. We varied the value of the initial proper time derivative of the sigma field expectation value and found that there is a narrow range of initial values that lead to the growth of instabilities, namely

$$0.25 fm^{-2} < |\dot\sigma| < 1.3 fm^{-2}.$$

Surprisingly when $|\dot\sigma| > 1.3 fm^{-2}$ instabilities no longer occur.

Figures 1-2 summarize the results of the numerical simulation for the evolution of the system (23),(28) and (32). We display the auxiliary field $\chi$ in units of $fm^{-2}$, the classical fields $\Phi$ in units of $fm^{-1}$ and the proper time in units of $fm^{-1}$. In Fig. 1 and Fig. 2 the proper time evolutions of the auxiliary field $\chi$ field are presented for two different values of $f_\pi$, where the initial conditions were fixed at $\tau = 1$. In the case of $f_\pi = 92.5 MeV$ we see that the instability lasts for less than $3 fm$. As discussed earlier, the regime of exponential growth of the unstable modes occurs whenever $\chi < 0$. We notice that when we rotate the expectation value from the $\sigma$ to the $\pi^1$ direction initially, the typical time that instability exists is of the same magnitude. We also notice that if we choose the time derivative to be zero and start from an equilibrium configuration, then the expansion alone is insufficient to generate instabilities. We find that in order to generate instabilities we require fluctuations in the classical kinetic terms such as $\dot\sigma$ or $\dot\pi_i$, and as discussed earlier, these initial conditions must be in a very narrow range to produce instabilities. Thus the rapid quench conditions assumed by other authors comprises only a small region of the phase space of initial fluctuations expected in an initial thermal distribution. As expected, at late proper times, the auxiliary field approaches its equilibrium value of $m_\pi^2$. For the case where $f_\pi = 125 MeV$, a value favored by fitting the low energy scattering, we see a very similar pattern, showing that our main results concerning the small range of initial conditions that lead to instabilities are not affected by a 30% change in the value of this parameter.

A crude estimate for the size of a disoriented chiral condensate is to multiply a typical life time of the instabilities by the speed of light. With this estimate the size of these regions are of the order of a few fermi. Of course one needs to study the growth of inhomogeneous instabilities to make any definite statements about this size. We were hoping that the correlation functions we have calculated could be interpreted easily in terms of a length parameter associated with the size of these regions. However, as we will see below, this didn't occur. In Fig. 3 we plot the proper time evolution of the classical fields $\sigma$ and $\pi$ for the same two choices for $f_\pi$ as in Fig. 1 and Fig. 2. We see that the sigma field asymptotes to its vacuum value $f_\pi$ and the $\pi$ field gradually converges to its equilibrium value of zero. In Fig. 4 we show the effect of starting the initial value problem at later times, namely $\tau_0 = 2.5, 4\, fm$, and compare them to the case previously studied with $\tau_0 = 1 fm$, which had a modest region of instability growth. We see that as we increase $\tau_0$ we decrease the possibility of instability growth, and with these late initial times it is not possible to produce instabilities even with kinetic energy fluctuations. In Fig. 5 we study the effect of the initial temperature on our time evolution problem. For $f_\pi = 92.5$ the critical temperature is $160 MeV$ (in the absence of explicit symmetry breaking term). We see that in the vicinity of the critical temperature the effects of varying the initial temperature is minor. In Fig. 6 we study the proper time evolution of the effective number density for various initial conditions. In thermal equilibrium one expects for an isentropic expansion in boost invariant coordinates that $s\tau = constant$. Since the number density is proportional to the entropy density one expects that once particle production stops that $n(\tau)\tau \to constant$. We see this trend in this figure, showing that we are reaching the out regime as the system expands. The breaks in some of these curves at early values of $\tau$ are a result of the fact that the interpolating number density cannot be defined when $\chi$ is negative. Next, we are interested in seeing how our results differ from the case where the system evolves in local thermal equilibrium. This is described by two correlation lengths, the inverse of the effective pion mass associated with $\chi(\tau)$, and the inverse of the proper time evolving effective temperature $T(\tau) = T(\tau_0)(\tau_0/\tau)^3$ corresponding to a boost invariant hydrodynamical models. We see from Fig. 7 that in the case of $\sigma(1) = \sigma_T$, $\pi^i(1) = 0$ and $\dot\sigma(1) = -1$, where maximum instability exists, complex structures are formed as contrasted to the local thermal equilibrium evolution. The interpolating phase space distribution $n(k_\eta, \mathbf{k}_\perp, \tau)$ obtained numerically, clearly exhibits a larger correlation length in the transverse direction than the $n_T(k_\eta, \mathbf{k}_\perp, \tau)$ equilibrium one, and has correlation in rapidity of the order of 1-2 units of rapidity. We notice that in both directions there are structures that does not lend themselves to a simple interpretation. On the other hand, the local thermal equilibrium evolution is quite regular apart from the normalization of the distributions that are changing with time due to oscillation in the quantity $\chi(\tau)$ which is damped to its equilibrium value once the system expands sufficiently. In Fig. 8 we look at the time evolution of the interpolating number operator for the case in which we did not have any fluctuations in the kinetic energy. Here we are still far from equilibrium, and we see that unlike the equilibrium case the correlation length in rapidity space is not decreasing with proper time. In this case where there are no instabilities we do not see complicated structures. The transverse distribution is similar to the equilibrium case. The pair density function $g(k_\eta, \mathbf{k}_\perp, \tau)$ is even more elusive to parametrize than the single particle distribution function. In the case where there are no instabilities ($\dot\sigma(1) = 0$) we see in Fig. 9 that although the transverse distribution is relatively simple, the distribution in the $\eta$ direction (whose fourier transform gives the rapidity distribution) has many length scales. When we have instabilities ($\dot\sigma(1) = -1$) then both distributions are complicated and possess several length scales, as seen in the lower part of Fig. 9. We also directly computed the fourier transform of the pion correlator (the first line of Eq.(40) for the case of $\pi_i = 0$, and as expected, it has even more complex structure than the distributions $n$ and $g$. In the case of non zero $\pi_i$ fields we expect that the correlator will have similar structure to the case described above, and it will involve at least 3 time evolving length scales related to the inverse mass of the pion, the inverse of the effective temperature and possible length scales describing domain growth.

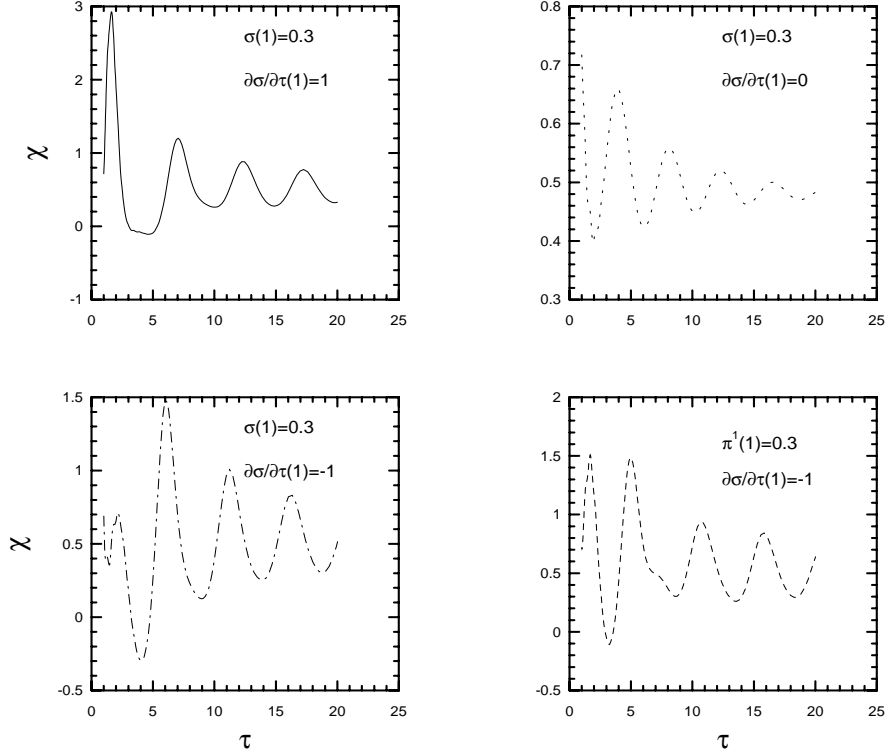

Figure 1: Proper time evolution of the $\chi$ field for four different initial conditions with $f_\pi = 92.5 MeV$.

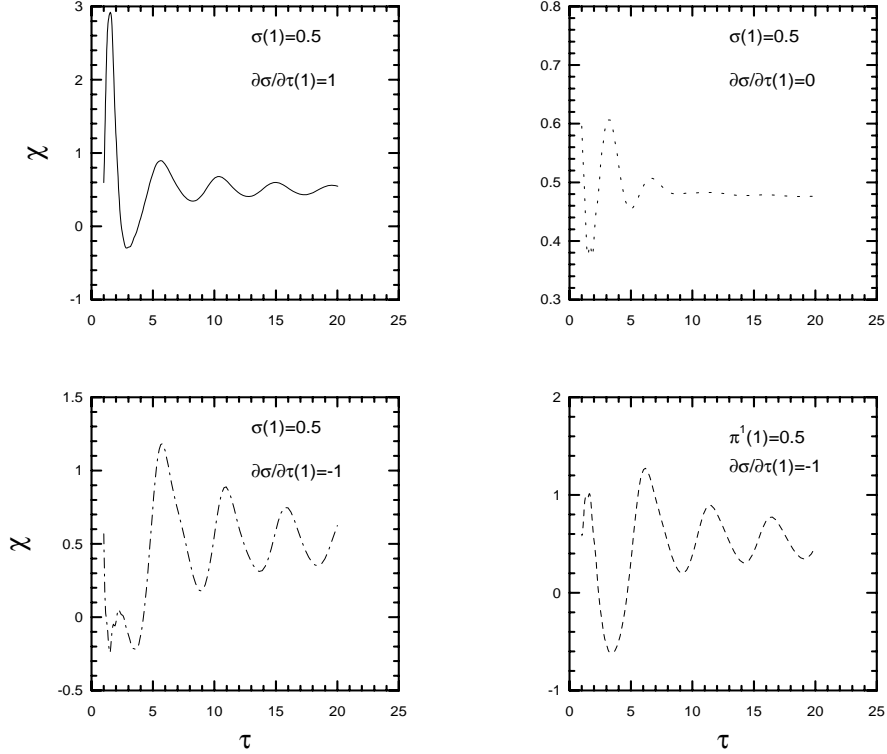

Figure 2: Same as Fig. 1, but for $f_\pi = 125 MeV$.

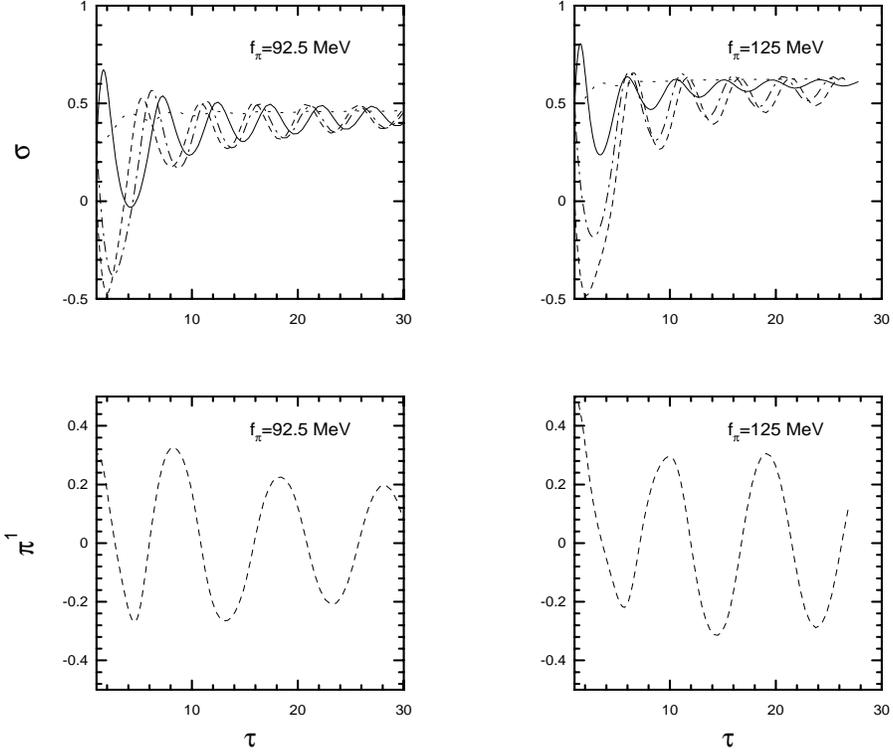

Figure 3: Proper time evolution of the $\sigma$ and $\pi$ fields for the following initial conditions: Solid line is for $\sigma(1) = \sigma_T$, $\pi^i(1) = 0$ and $\dot{\sigma}(1) = 1$. Dotted line is for $\sigma(1) = \sigma_T$, $\pi^i(1) = 0$ and $\dot{\sigma}(1) = 0$. Dashed- dotted line is for $\sigma(1) = \sigma_T$, $\pi^i(1) = 0$ and $\dot{\sigma}(1) = -1$. Dashed line is for $\sigma(1) = 0$, $\pi^i(1) = \sigma_T$ and $\dot{\sigma}(1) = -1$. At $T = 200 MeV$, $\sigma_T = 0.3 fm^{-1}$ for $f_\pi = 92.5 MeV$ and $\sigma_T = 0.5 fm^{-1}$ for $f_\pi = 92.5 MeV$.

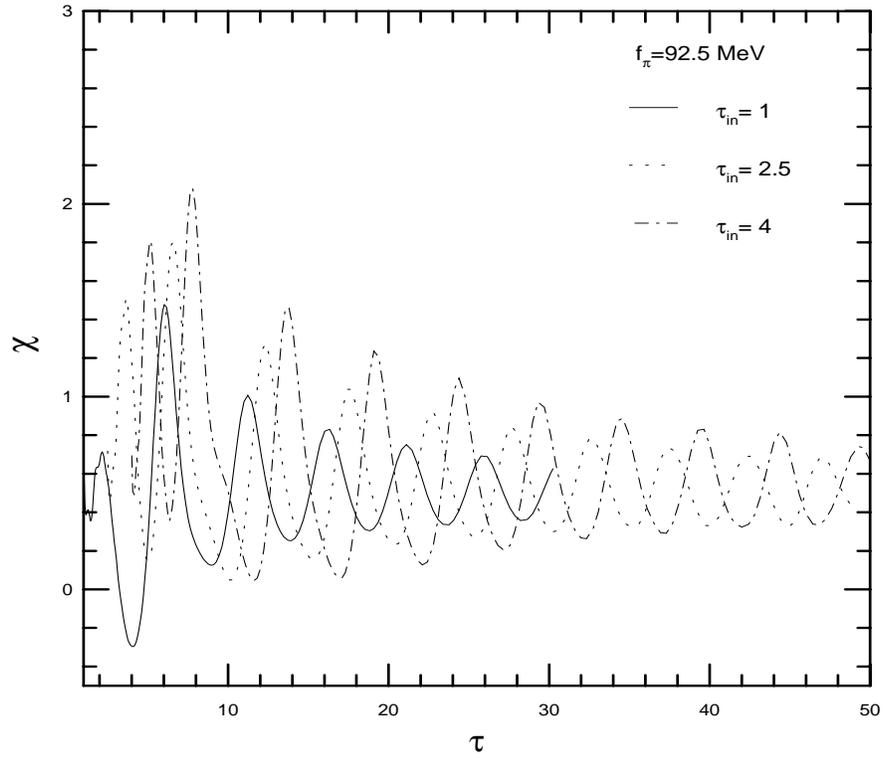

Figure 4: Proper time evolution of the $\chi$ field for three different initial proper times $\tau_0 = 1, 2.5, 4$ with $f_\pi = 92.5 MeV$, $\sigma(\tau_{in}) = \sigma_T$, $\pi^i(\tau_{in}) = 0$ and $\dot\sigma(\tau_{in}) = -1$.

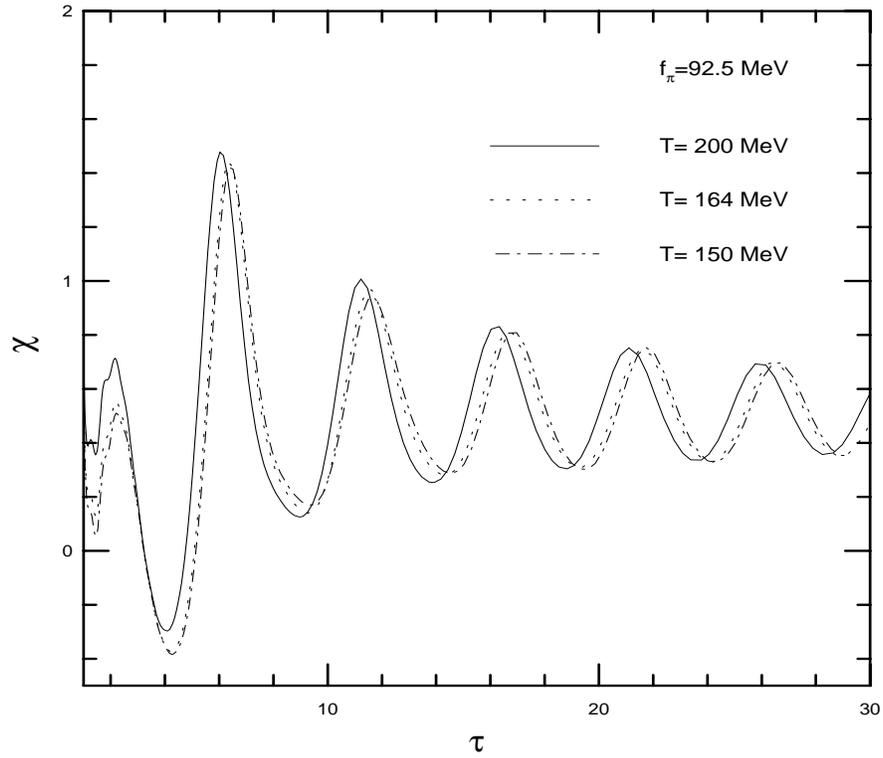

Figure 5: Proper time evolution of the $\chi$ field for three different initial thermal distributions with $T = 200, 164, 150 MeV$ for the case of $\sigma(1) = \sigma_T$, $\pi^i(1) = 0$ and $\dot{\sigma}(1) = -1$.

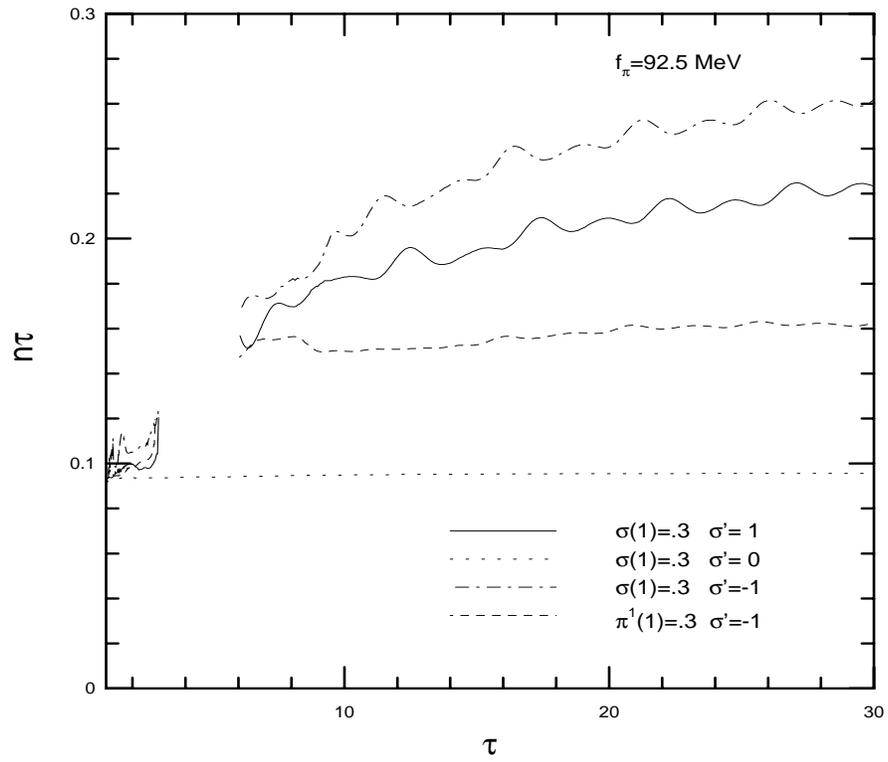

Figure 6: Proper time evolution of the produced particle density $n\tau \equiv dN/d\eta dx_\perp$ for the same evolution shown in Fig. 1.

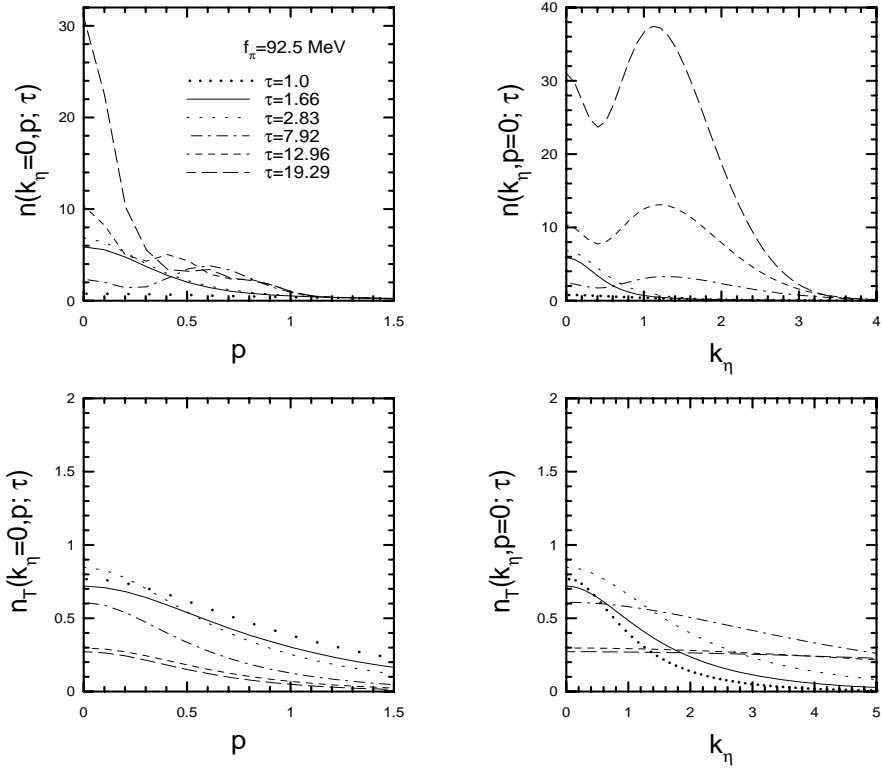

Figure 7: Slices of $k_\eta = 0$ and $p \equiv |\mathbf{k}_\perp| = 0$ of the proper time evolution of the interpolating phase space particle number density $n(k_\eta, \mathbf{k}_\perp, \tau)$ for $\sigma(1) = \sigma_T$, $\pi^i(1) = 0$ and $\dot\sigma(1) = -1$ compared with the corresponding local thermal equilibrium densities $n_T(k_\eta, \mathbf{k}_\perp, \tau)$.

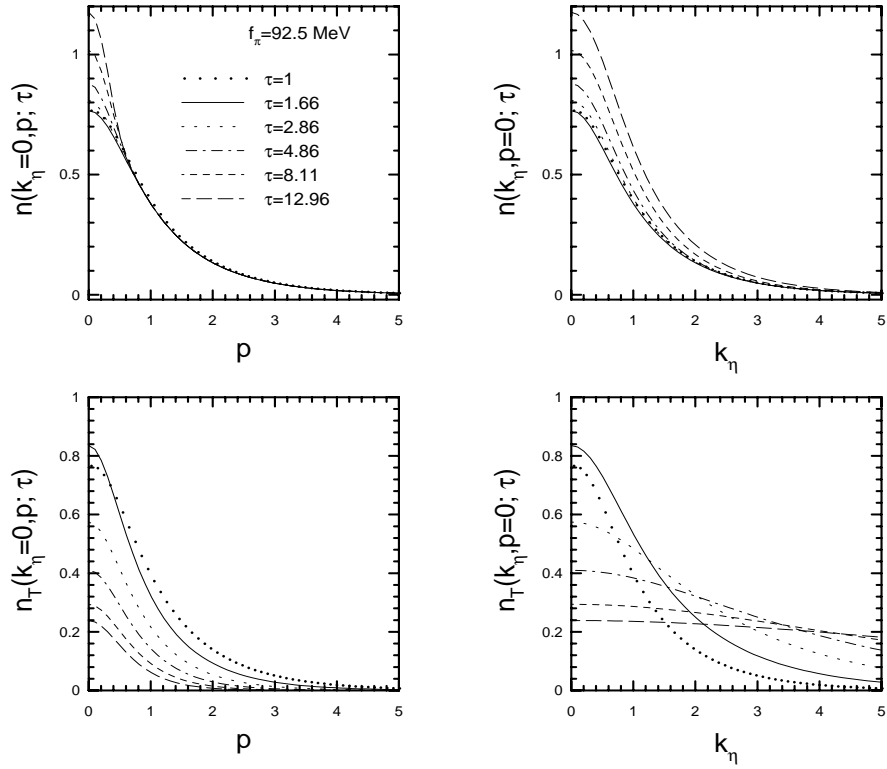

Figure 8: As in Fig. 7, but for $\sigma(1) = \sigma_T$, $\pi^i(1) = 0$ and $\dot\sigma(1) = 0$.

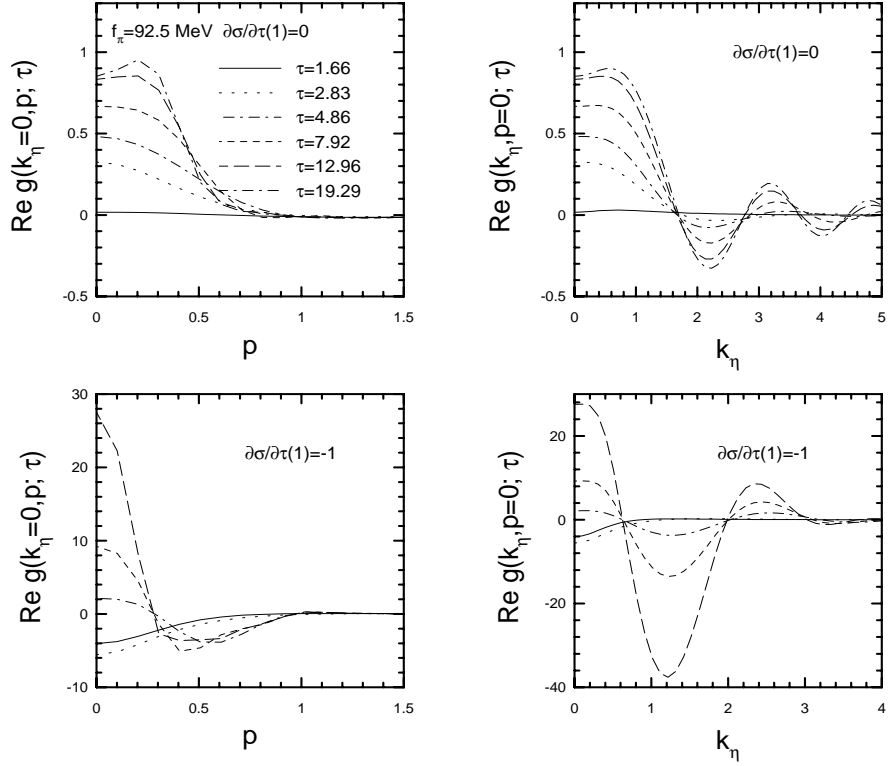

Figure 9: Slices of $k_\eta = 0$ and $p \equiv |\mathbf{k}_\perp| = 0$ of the proper time evolution of the real part of the phase space pair density $g(k_\eta, \mathbf{k}_\perp, \tau)$ for $\sigma(1) = \sigma_T$, $\pi^i(1) = 0$ and $\dot\sigma(1) = 0, -1$.

## 5. Conclusions

In this paper we have performed numerical simulations in the regime of the chiral phase transition in the linear sigma model for a wide variety of initial conditions starting above the critical temperature for an expanding plasma of pions and sigmas. Assuming that this model gives a reasonable description in this temperature range, we found initial conditions where instabilities grew in the scenario required for the formation of disoriented chiral condensates. The constraints on our model which are imposed in order to approximate the low energy physics of pion interactions require a large renormalized coupling constant. This had the effect of rapidly damping the instabilities, and we found no evidence in this model for large domains of disoriented chiral condensates. We did, however, see rather large departures in the phase space number density from one which would result from an evolution in local thermal equilibrium. These departures show a narrowing of the momentum space distribution which could be interpreted as a larger spatial correlation length. However we did not find a simple way of extracting correlation lengths from our results. In our simulations we assumed a reasonable mechanism for cooling– namely the expansion of the plasma following its production in a collision. In the future we hope to include scattering effects which will introduce another time scale, namely the equilibration time scale, into the problem, as well as study the nonlinear sigma model to see if the results differ significantly from those found here.

## 6. Acknowledgements

This work was done in collaboration with Juan Pablo Paz, Fred Cooper and Emil Mottola. I would like to thank Alex Kovner for the many enlightening discussions we had, which were very valuable and Salman Habib for stimulating criticisms and discussions. This research was performed in part using the resources located at the Advanced Computing Laboratory of Los Alamos National Laboratory, Los Alamos, NM 87545.